\def\be{\begin{equation}}
\def\ee{\end{equation}}
\def\ba{\begin{array}}
\def\ea{\end{array}}

\def\Rb{{I\!\! R}}
\def\Cb{{\Bbb C}}
\documentstyle[12pt]{article}
\begin{document}
\input amssym.def
\centerline{\Large\bf A Note on Pseudo-Hermitian Systems with Point}
\medskip
\centerline{\Large\bf Interactions and Quantum Separability}
\bigskip

\centerline{Shao-Ming Fei}
\bigskip

\centerline{Department of Mathematics, Capital Normal University, Beijing 100037}
\centerline{Institute of Applied Mathematics, University of Bonn, D-53115 Bonn}

\bigskip
\medskip
\bigskip
\medskip

\centerline{\large Abstract}
\vspace{2ex}

We study the quantum entanglement and separability of
Hermitian and pseudo-Hermitian
systems of identical bosonic or fermionic particles with point interactions.
The separability conditions are investigated in detail.
\bigskip
\medskip

{\raggedleft PACS numbers: {02.30.Ik, 11.30.Er, 03.65.Fd}}

{\raggedleft Keywords: Point interactions, PT-symmetry, Separability}
\bigskip
\bigskip
\bigskip

Pseudo-Hermitian quantum mechanical systems have been extensively
investigated due to some mathematical and physical considerations \cite{pt}.
For PT-symmetric systems with singular contact interactions, we have
studied the classification, spectra
and integrability problem in \cite{pt-fei}.
The results are generalized to the case of identical
bosonic and fermionic many-body systems with PT-symmetric contact interactions
and spin-coupling interactions \cite{pt-spin}.

On the other hand, as one of the most striking features of quantum phenomena
\cite{q}, quantum entanglement has been identified as a key
non-local resource in quantum information processing such as
quantum computation \cite{book}, quantum teleportation,
dense coding, quantum cryptographic
schemes, entanglement swapping and remote
state preparation (RSP).
Nevertheless the theory of quantum entanglement is far from being
satisfied. One of main problem in quantum entanglement is to
judge a quantum state is separable or entangled, i.e. the separability.
In this paper we study separability of
Hermitian and pseudo-Hermitian
systems of many body systems with contact interactions.

We first consider contact interactions without spin coupling.
A particle moving in one dimension with
point interactions at the origin can be characterized by the
non-separated\footnote{The word ``separated" here in point
interaction is different from
the ``separability" in quantum entanglement.}
boundary conditions imposed on the wave
function $\varphi$ at $x=0$,
\be\label{BOUND}
\left( \begin{array}{c}
\varphi\\
\varphi'\end{array} \right)_{0^+}
=\left(
\begin{array}{cc}
\textsc{a} & \textsc{b} \\
\textsc{c} & \textsc{d} \end{array} \right)\left( \begin{array}{c}
\varphi\\
\varphi'\end{array} \right)_{0^-},
\end{equation}
where $\textsc{a}$, $\textsc{b}$, $\textsc{c}$, $\textsc{d}$ are some
complex numbers subject to certain conditions. For instance,
if $(\textsc{a,b,c,d})=e^{i\theta}(a,b,c,d)$, with $ad-bc = 1$, $\theta, a,b,c,d \in \Rb$,
the system is self-adjoint. If $(\textsc{b,c})=e^{i\theta}(b,c)$,
$\textsc{a}=e^{i\theta}\sqrt{1 +bc}\,e^{i\phi}$,
$\textsc{d}=e^{i\theta}\sqrt{1 +bc}\,e^{-i\phi}$,
with the real parameters $ b \geq 0$, $c \geq -1/b$ (if $b\neq 0$),
$\theta, \phi \in [0, 2 \pi)$, then the system is
PT-symmetric.

Another kind of point interaction is described by so called
separated boundary conditions of the form,
\be\label{bounds}
\varphi^\prime(0_+) = h^+ \varphi (0_+)~, ~~~
\varphi^\prime(0_-) = h^- \varphi (0_-).
\ee
It represents self-adjoint systems for
$h^{\pm} \in \Rb \cup \{ \infty\}$, or
$ h^+ =h^- = \infty$ (Dirichlet boundary conditions), or
$h^+ = h^- = 0$ (Neumann boundary conditions).
For $h^+ = h_1 e^{i\theta}$, $h^- =  - h_1 e^{-i\theta}$,
with the real $h_1$ and phase parameter $ \theta \in [0,2\pi)$
the system is PT-symmetric.

We consider now the quantum entanglement of many body systems
with contact interactions described by the boundary conditions
(\ref{BOUND}) and (\ref{bounds}), when any two particles
meet together. We first consider two-particle case.
For a particle with spin $s$, the wave function has $n=2s+1$ components.
Let ${\cal H}$ denote the spin vector space with basis
$e_\alpha$, $\alpha=1,...,n$. The wave function of
a two-particle system is an $n^2$-dimensional column vector
in ${\cal H}\otimes{\cal H}$ and can be generally expressed as
\be\label{psia}
\psi=\sum_{\alpha,\beta=1}^n\phi_{\alpha
\beta}e_\alpha\otimes e_\beta.
\ee
The state $\psi$ is separable (not entangled) in the spin space
if it can be written as a product vector,
$$
\sum_{\alpha=1}^n\xi_{\alpha}e_\alpha\otimes \sum_{\beta=1}^n \eta_{\beta} e_\beta
$$
for some $\xi_\alpha$, $\eta_\beta\in\Cb$. It is shown that $\psi$ is separable
if and only if the concurrence
\be\label{cn}
C=\sqrt{\displaystyle
\frac{n}{2(n-1)}\sum_{\alpha,\beta,\gamma,\delta=1}^n \vert
\phi_{\alpha \gamma}\phi_{\beta \delta}-\phi_{\alpha \delta}\phi_{\beta \gamma}\vert^2}
\ee
is zero\cite{UR,entangle1}.

Let $x_1$, $x_2$ be the coordinates and  $k_1$, $k_2$ the momenta
of the two particles respectively.
Each particle has $n$-`spin' states designated by $s_1$ and $s_2$,
$1\leq s_i\leq n$.
For $x_1\neq x_2$, these two particles are free. The
wave functions $\psi$ are symmetric (resp. antisymmetric) with respect
to the interchange $(x_1,s_1)\leftrightarrow(x_2,s_2)$
if $s$ is an integer (resp. half integer).
In the region $x_1<x_2$, in terms of Bethe ansatz the
wave function has the following form
\be\label{w1}
\psi=u_{12}e^{i(k_1x_1+k_2x_2)}+u_{21}e^{i(k_2x_1+k_1x_2)}
=u_{12}e^{i(K_{12} X-k_{12}x)}+u_{21}e^{i(K_{12}X+k_{12}x)},
\ee
where $X=(x_1+x_2)/2$, $x=x_2-x_1$ are the coordinates of the
center of mass system, $K_{12}=k_1+k_2$, $k_{12}=(k_1-k_2)/2$,
$u_{12}$ and $u_{21}$ are $n^2$-dimensional column vectors, representing
the spin part of the wave function.

In the region $x_1>x_2$,
\be\label{w2}
\psi=(P^{12}u_{12})e^{i(K_{12} X+k_{12}x)}
+(P^{12}u_{21})e^{i(K_{12} X-k_{12}x)},
\ee
where according to the symmetry or antisymmetry conditions,
$P^{12}={\bf p}^{12}$ for bosons and $P^{12}=-{\bf p}^{12}$ for fermions, ${\bf p}^{12}$
being the operator on the $n^2$-dimensional column vectors that interchanges
the spins of the two particles.
Substituting (\ref{w1}) and (\ref{w2}) into the boundary
conditions (\ref{BOUND}) at $x=0$, we get
\be\label{a1}
\left\{
\begin{array}{l}
u_{12}+u_{21}
=\textsc{a} P^{12}(u_{12}+u_{21})+
i \textsc{b} k_{12}P^{12}(u_{12}-u_{21}),\\
ik_{12}(u_{21}-u_{12})
=\textsc{c} P^{12}(u_{12}+u_{21})+i\textsc{d} k_{12}P^{12}(u_{12}-u_{21}).
\end{array}\right.
\ee
Eliminating the term $P^{12}u_{12}$ from (\ref{a1})
we obtain the relation
\be\label{2112}
u_{21} = Y_{21}^{12} u_{12}~,
\ee
where
\be\label{a21a12}
Y_{21}^{12}
=\frac{
2ik_{12}(\textsc{a}\textsc{d}-\textsc{b}\textsc{c})P^{12}+ik_{12}
(\textsc{a}-\textsc{d})+(k_{12})^2\textsc{b}+\textsc{c}}
{ik_{12}(\textsc{a}+\textsc{d}) + (k_{12})^2\textsc{b}-\textsc{c}}.
\ee

Similarly from the interaction given by the separated boundary condition (\ref{bounds})
we have $h^+ = - h^-\equiv h \in \Rb \cup \{ \infty\}$ and
\be\label{a21a12s}
Y_{21}^{12}
=\frac{ik_{12} + h}{ik_{12} - h} ~.
\ee

Comparing (\ref{w1}) with (\ref{psia}) and using (\ref{2112}) we have
\be\label{phiab}
\phi_{\alpha\beta}=(u_{12})_{\alpha\beta}e^{i(k_1x_1+k_2x_2)}+(Y_{21}^{12}u_{12})_{\alpha\beta}e^{i(k_2x_1+k_1x_2)}.
\ee
From (\ref{cn}), state (\ref{w1}) is separable if and only if
$\vert\phi_{\alpha \gamma}\phi_{\beta \delta}-\phi_{\alpha \delta}\phi_{\beta \gamma}\vert=0$,
$\alpha$, $\gamma$, $\beta$, $\delta=1,...,n$. We have, for general $x_1$, $x_2$, $k_1$, $k_2$,
\be\label{c1}
(u_{12})_{\alpha\gamma}(u_{12})_{\beta\delta}-
(u_{12})_{\alpha\delta}(u_{12})_{\beta\gamma}=0,
\ee
\be\label{c2}
(Y_{21}^{12}u_{12})_{\alpha\gamma}(Y_{21}^{12}u_{12})_{\beta\delta}
-(Y_{21}^{12}u_{12})_{\alpha\delta}(Y_{21}^{12}u_{12})_{\beta\gamma}=0,
\ee
\be\label{c3}
\ba{l}
(u_{12})_{\alpha\gamma}(Y_{21}^{12}u_{12})_{\beta\delta}
-(u_{12})_{\alpha\delta}(Y_{21}^{12}u_{12})_{\beta\gamma}\\[2mm]
~~~~~~~+(Y_{21}^{12}u_{12})_{\alpha\gamma}(u_{12})_{\beta\delta}
-(Y_{21}^{12}u_{12})_{\alpha\delta}(u_{12})_{\beta\gamma}=0.
\ea
\ee

The conditions (\ref{c1}) stands for that the vector $u_{12}$ itself
should be separable.
For the interactions given by the separated boundary condition (\ref{bounds})
characterized by the operator (\ref{a21a12s}), we have that
if $u_{12}$ is separable, the vector $Y_{21}^{12}u_{12}$ is also separable,
and eq. (\ref{c3}) is satisfied as well. Hence in this case
the wave function (\ref{psia}) of the spin part is separable if and only if
$u_{12}$ is separable.

Nevertheless, associated with the boundary condition (\ref{BOUND}), the operator (\ref{a21a12})
is not an identity one. Noting that $(P^{12}u_{12})_{\alpha\beta}=
(u_{12})_{\beta\alpha}$ for bosons and
$(P^{12}u_{12})_{\alpha\beta}=-(u_{12})_{\beta\alpha}$ for fermions,
we still have that the vector $Y_{21}^{12}u_{12}$ is also separable
if $u_{12}$ is separable. But condition (\ref{c3}) is satisfied
only when $(P^{12}u_{12})_{\alpha\beta}=(u_{12})_{\beta\alpha}$.
Therefore in this case the systems is separable if and only if
$u_{12}$ is separable, and $(u_{12})_{\alpha\beta}=(u_{12})_{\beta\alpha}$
for bosons and $(u_{12})_{\alpha\beta}=-(u_{12})_{\beta\alpha}$ for fermions.

For $N$-body ($N\geq 3$) systems, in the region $x_1<x_2<...<x_N$, the wave function is given by
\be\label{psiN}
\ba{rcl}
\Psi&=&\displaystyle\sum_{\alpha_1,..,\alpha_N=1}^n
\phi_{\alpha_1,...,\alpha_N}(x_1,...,x_N)e_{\alpha_1}\otimes...\otimes e_{\alpha_N}\\[5mm]
&=&\displaystyle u_{12...N}e^{i(k_1x_1+k_2x_2+...+k_Nx_N)}
+u_{21...N}e^{i(k_2x_1+k_1x_2+...+k_Nx_N)}\\[2mm]
&&+(N!-2)~other~terms.
\ea
\ee
The column vectors $u_{12...N}$ now have $n^N$ dimensions. Other vectors $u$
are obtained from the following relations,
\be\label{a1n}
u_{\alpha_1\alpha_2...\alpha_i\alpha_{i+1}...\alpha_N}=Y_{\alpha_{i+1}\alpha_i}^{ii+1}
u_{\alpha_1\alpha_2...\alpha_{i+1}\alpha_i...\alpha_N},
\ee
where
\be\label{y}
Y_{\alpha_{i+1}\alpha_i}^{ii+1}=
\frac{2ik_{\alpha_i \alpha_{i+1}}(\textsc{a}\textsc{d}-\textsc{b}\textsc{c})P^{ii+1}
+ik_{\alpha_i\alpha_{i+1}}(\textsc{a}-\textsc{d}) + (k_{\alpha_i\alpha_{i+1}})^2 \textsc{b}+\textsc{c}}
{ik_{\alpha_i\alpha_{i+1}}(\textsc{a}+\textsc{d})+(k_{\alpha_i\alpha_{i+1}})^2 \textsc{b}-\textsc{c}}
\ee
for non-separated boundary conditions, and
\be\label{ys}
Y_{\alpha_{i+1}\alpha_i}^{ii+1}=
\frac{ik_{\alpha_i\alpha_{i+1}} + h}{ik_{\alpha_i\alpha_{i+1}} - h}
\ee
for separated boundary condition.
Here $k_{\alpha_i\alpha_{i+1}}=(k_{\alpha_i}-k_{\alpha_{i+1}})/2$,
$P^{ii+1}={\bf p}^{ii+1}$ for bosons and $P^{ii+1}=-{\bf p}^{ii+1}$ for fermions,
with ${\bf p}^{ii+1}$ the operator on the $n^N$-dimensional column vectors
that interchanges $s_i\leftrightarrow s_{i+1}$.
Due to the Yang-Baxter equation\cite{yb}, $\textsc{a}$, $\textsc{b}$, $\textsc{c}$ and $\textsc{d}$
have to be real and satisfy $\textsc{b}=0$, $\textsc{a}=\textsc{d}=\pm 1$,
which is the case of both self-adjoint and PT-symmetric interactions.

The state $\Psi$ is separable if and only if the generalized concurrence \cite{entangle1}
\be\label{cnm}
C_N=\sqrt{\displaystyle\frac{N}{2(N-1)(2^{M-1}-1)}\sum_p
\sum_{\{\alpha,\alpha^\prime,\beta,\beta^\prime\}}^N
\vert \phi_{\alpha\beta}\phi_{\alpha^\prime\beta^\prime}-
\phi_{\alpha\beta^\prime}\phi_{\alpha^\prime\beta}\vert^2}=0\,,
\ee
where $\alpha$ and $\beta$, as well as $\alpha^\prime$ and $\beta^\prime$
are two subsets of the sub-index of $\phi$ in (\ref{psiN}), such that they together span
the whole index space, $\alpha$ and $\alpha^\prime$, as well as $\beta$ and $\beta^\prime$
have the same number of indices, $\displaystyle\sum_p$ stands for the summation
over all possible combinations of the indices of $\alpha$, $\beta$,
$\alpha^\prime$ and  $\beta^\prime$.
From (\ref{psiN}) we have
$$
\ba{rcl}
\phi_{\alpha_1,...,\alpha_N}
&=&(u_{12...N})_{\alpha_1,...,\alpha_N}e^{i(k_1x_1+k_2x_2+...+k_Nx_N)}\\[3mm]
&&+(Y_{21}^{12}u_{12...N})_{\alpha_1,...,\alpha_N}e^{i(k_2x_1+k_1x_2+...+k_Nx_N)}
+(N!-2)~terms.
\ea
$$
For contact interactions associated with the boundary condition
(\ref{bounds}), $C_N=0$ is satisfied if $u_{12...N}$ is separable.
For those associated with the boundary condition (\ref{BOUND}),
the systems is separable if and only if

i) $u_{12...N}$ is separable;

ii) $p^{ij}u_{12...N}=u_{12...N}$ for bosons, $i\neq j=1,...,N$;

iii) $p^{ij}u_{12...N}=-u_{12...N}$ for fermions, $i\neq j=1,...,N$.

We consider now two particles with contact interactions and spin-coupling
interactions. Corresponding to the non-separated case, a general boundary
condition described in the center of mass coordinate system has a form:
\be\label{BOUND1}
\left( \begin{array}{c}
\varphi\\
\varphi'\end{array} \right)_{0^+}
=\left(
\begin{array}{cc}
A & B \\
C & D \end{array} \right)
\left( \begin{array}{c}
\varphi\\
\varphi'\end{array} \right)_{0^-},
\end{equation}
where $\varphi$ and $\varphi'$ are $n^2$-dimensional
column vectors, $A,B,C$ and $D$ are
$n^2\times n^2$ matrices.
Corresponding to the separated boundary condition
(\ref{bounds}) one has
\be\label{bounds1}
\varphi^\prime(0_+) = F \varphi (0_+), ~~~
\varphi^\prime(0_-) = G \varphi (0_-) .
\ee
For the self-adjoint point interactions, $G$ and $F$ are just
$n^2\times n^2$ Hermitian matrices.
For PT-symmetric interactions, $G=-F^{\ast}$.
The integrability and separability for general
case are quite complicated. To elucidate the problem we
consider the separated boundary condition (\ref{bounds1})
in the following. For a many-body system with contact
interactions described by (\ref{bounds1}),
the wave function (\ref{psiN}) is no longer separable
if and only if $u_{12...N}$ is separable.

In stead of (\ref{ys}), we have
\be\label{ys1}
Y_{\alpha_{j+1}\alpha_j}^{jj+1}=[ik_{\alpha_j\alpha_{j+1}}-F_{j{j+1}}]^{-1}
[ik_{\alpha_j\alpha_{j+1}} + F_{jj+1}],
\ee
where $F_{jj+1}$ stands for the application of the operator $F$ to the $j$-th and $(j+1)$-th particles.
$F$ is a real $n^2\times n^2$ matrix that commutes with the permutation operator ${\bf p}$.
For the case of spin $s=1/2$ (i.e. $n=2$), $F$ is generally of the form
\be\label{hspin}
F=\left(\ba{cccc}a&e_1&e_1&c\\
e_3&f&g&e_2\\e_3&g&f&e_2\\
d&e_4&e_4&b\ea\right),
\ee
where $a,b,c,d,f,g,e_1,e_2,e_3,e_4\in\Rb$. By straightforward calculations one
can show that if $e_i=0$, $i=1,...,4$, $a=b=f+g$, then the state $(\ref{psiN})$
is separable ($C_N=0$) if and only if conditions i)-iii) above are satisfied.

For general spin values, we can show that $(\ref{psiN})$ could be separable if
$F=f + g {\bf p}$, $f,g\in\Rb$, where $f$ is understood as multiplied by an
$n^2\times n^2$ identity matrix. Since in this case, $Y_{\alpha_{j+1}\alpha_j}^{jj+1}$
in (\ref{ys1}) has the form
\be\label{ysF}
Y_{\alpha_{j+1}\alpha_j}^{jj+1}=\frac{-g(g^2-f^2-k_{\alpha_j\alpha_{j+1}}^2+
2igk_{\alpha_j\alpha_{j+1}}{\bf p}^{jj+1})}{(ik_{\alpha_j\alpha_{j+1}}-f)^2-g^2}.
\ee
Therefore as long as the conditions i)-iii) are satisfied,
$(\ref{psiN})$ is separable. Here the contact interaction system described
by the operator $F=f + g {\bf p}$ is both self-adjoint and PT-symmetric.

We have studied the quantum separability of
identical bosonic or fermionic particles with point interactions.
The separability conditions depend on the $Y$-operators.
For the separated boundary condition (\ref{bounds}), the $Y$ operator
(\ref{ys}) is just an identity one. Hence the many-body state is separable
if one of the $u$ vectors in (\ref{psiN}) is separable.
For general no-separated boundary condition (\ref{BOUND})
or separated boundary condition with spin-couplings (\ref{bounds1}),
the $Y$ operators (\ref{y}), (\ref{ysF}) are proportional to
the permutation operator ${\bf p}$. (\ref{psiN}) is then separable
if the conditions i)-iii) are satisfied. One may also study the
separability of systems characterized by the boundary condition
(\ref{BOUND1}) as well, according to detailed $A$, $B$, $C$, $D$
for the requirement of self-adjointness or PT-symmetry, though
it could be quite complicated.

\end{document}